\documentclass[fleqn,10pt]{wlscirep}
\usepackage[utf8]{inputenc}
\usepackage[T1]{fontenc}
\usepackage{changes}
\usepackage{figcaps}
\usepackage[most]{tcolorbox}

\usepackage[wby]{callouts}
\usepackage{graphicx}
\usepackage[export]{adjustbox}
\usepackage{multicol}
\usepackage[english]{babel}

\usepackage{wrapfig}
\usepackage{amsmath,fixmath}
\usepackage{lineno}
\usepackage[explicit]{titlesec}
\usepackage{hyperref}

\usepackage[nameinlink]{cleveref}
\Crefformat{app}{Appendix #1}
\Crefformat{sec}{Introduction #1}

\Crefname{section}{Sec.}{Secs.}
\Crefname{figure}{Fig.}{Figs.}
\Crefname{equation}{Eq.}{Eqs.}

\usepackage{graphicx}
\usepackage[labelfont=bf]{caption}
\usepackage[format=hang]{subcaption}

\usepackage{tabularx}
\usepackage[justification=justified, format=plain]{caption}

\usepackage{algorithm,algorithmic}

\hypersetup{
 bookmarks=true, 
 unicode=false, 
 pdftoolbar=true, 
 pdfmenubar=true, 
 pdffitwindow=false, 
 pdfstartview={FitH}, 
 pdftitle={}, 
 pdfauthor={}, 
 pdfcreator={}, 
 pdfproducer={}, 
 pdfkeywords={} {} {}, 
 pdfnewwindow=true, 
 colorlinks=true, 
 linkcolor=blue, 
 filecolor=magenta, 
 urlcolor=cyan 
}

\usepackage{tikz}
\usetikzlibrary{shapes.arrows}
\usetikzlibrary{arrows.meta}

\usepackage{tikz,caption}
\usepackage{hyperref}
\usetikzlibrary{calc}
\DeclareCaptionFormat{overlay}{\gdef\capoverlay{#1#2#3\par}}
\DeclareCaptionStyle{overlay}{format=overlay}

\newcommand{\be}{\begin{equation}}
\newcommand{\ee}{\end{equation}} 
\newcommand{\bea}{\begin{eqnarray}}
\newcommand{\eea}{\end{eqnarray}}

\renewcommand{\v}[1]{\mathbf{#1}} 

\newcommand{\ccup}[1]{\left\{#1\right\}}

\renewcommand{\t}[1]{\tilde{#1}}

\renewcommand{\ref}[1]{[\ref{#1}]}

\makeatletter
\renewcommand\@biblabel[1]{[#1]}
\makeatother
\graphicspath{{./figures/}} 

\usepackage{enumitem,amssymb}
\newlist{todolist}{itemize}{2}
\setlist[todolist]{label=$\square$}
\newlist{todolist_done}{itemize}{2}
\setlist[todolist_done]{label=$\blacksquare$}
\usepackage{pifont}
\usepackage[normalem]{ulem} 

\newcommand{\nextrout}{\mbox{{\small Nextrout}}}
\newcommand{\pp}{\textit{Physarum polycephalum}}
\newcommand{\source}{origin}
\newcommand{\sink}{destination}

\begin{document}

\title{Urban transportation networks and optimal transport-based infrastructures: similarity and economy of scale} 

\author[1,*]{Daniela Leite}
\author[1,*]{Caterina De Bacco}
\affil[1]{Max Planck Institute for Intelligent Systems, Cyber Valley, 72076, Tübingen, Germany}

\affil[*]{caterina.debacco@tuebingen.mpg.de}
\figcapsoff
\printfigures

\begin{abstract}
	Designing and optimizing the structure of urban transportation networks is a challenging task. In this study, we propose a method inspired by optimal transport theory and the principle of economy of scale that uses little information in input to generate structures that are similar to those of public transportation networks. Contrarily to standard approaches, it does not assume any initial backbone network infrastructure but rather extracts this directly from a continuous space using only a few origin and destination points. Analyzing a set of urban train, tram and subway networks, we find a noteworthy degree of similarity in several of the studied cases between simulated and real infrastructures. By tuning one parameter, our method can simulate a range of different subway, tram and train networks that can be further used to suggest possible improvements in terms of relevant transportation properties. Outputs of our algorithm provide naturally a principled quantitative measure of similarity between two networks that can be used to automatize the selection of similar simulated networks.

\end{abstract}

\thispagestyle{empty}
\flushbottom
\maketitle

\section*{INTRODUCTION} \label{sec:introduction}

Transportation networks are a fundamental part of a city's infrastructure. Their design impacts the efficiency with which the system is operated, hence, they should follow optimal principles while being constrained by limitations like budget or physical obstacles. Existing approaches for studying the quality of network design often rely on the analysis of the topological network properties, and relate them to optimal features like transportation cost, efficiency or robustness. These analyses are usually made \textit{a posteriori}, only once the network has been constructed, and thus only resulting properties can be analyzed \cite{zhang2013networked,ding2019application}. A different approach is that of posing \textit{a priori} a principled optimization setup, where one defines a cost function that a network should minimize under a set of constraints, and then searches for optimal solutions in terms of network topologies.
Numerous studies have explored this approach in biological networks, transportation networks, etc. \cite{navlakha2011algorithms,easley2010networks}. However, most of these methods rely on an existing backbone of a network infrastructure that can be optimized in terms of traffic distribution \cite{watanabe2011traffic,liu2013physarum} but do not consider the possibility of building the network from scratch, starting from a limited set of nodes. Alternatively, as optimizing over all possible
topologies is difficult, one can investigate only various simple shapes from a predetermined set of possible geometries \cite{aldous2019optimal,mc2020role,barthelemy2022optimal} or rely on heuristics \cite{cantarella2006heuristics,laporte2011planning}. Recently, Kay et al.\cite{kay2022stepwise} presented a two-step agent-based model that replicates biologically-grown networks and proposes them as a template for urban design.
Further, the lack of a principled metric to measure the similarities between an observed network and a simulated one poses the problem of making this evaluation effectively.

In this work, we show that urban transportation systems can exhibit underlying
network topologies similar to those that follow optimality principles as defined in optimal transport theory. Specifically, we
propose a model to characterize real transportation networks based on a simple optimal transport framework, similar to what is observed in biological systems 
like the slime mold \pp, which adapts its network structure to reach food patches in an optimal way.
A previous study by Tero et al.\cite{tero2010rules} shows how this mold forms networks with comparable optimal transportation properties, e.g. efficiency and cost, to those of the Tokyo rail system, but provided no rigorous quantitative definition of network similarity beyond measuring these properties. 
We empirically validate our approach with a systematic characterization of the structure of several urban transportation networks and propose a rigorous definition of similarity in terms of optimal transport theory.

Urban transportation networks often exhibit different network structures based on the goals of network designers.
For instance, some networks focus on connecting people living in the outer layers of the city to the city core, while others
prefer to develop a robust infrastructure servicing the core \cite{derrible2010characterizing}. 
Several studies have focused on analyzing properties like scaling laws and network connectivity
\cite{levinson2012network,derrible2010complexity,lin2013complex}, which are indications of an underlying optimality mechanism that these networks might follow to make a city efficient, both in reduced infrastructure costs per capita and in increased productivity. 
However, our understanding of what optimality principles are captured in real transportation networks is incomplete. In fact,  studying network properties could only partially explain the underlying mechanisms regulating network design, as each property captures a different aspect. 
Here, we take a different approach and build the network from scratch while comparing it with the real ones observed from data, starting with only a few shared nodes in input. Specifically, we model network structures observed in urban transportation networks by adapting a classical optimal transport framework to simulate a network-design problem dependent on realistic travel demand settings and using little information in input.  We then compare the resulting networks with those observed from real data and assess their similarity. Importantly, the model can simulate different optimal strategies by tuning a parameter $\beta$,  which interpolates between minimizing infrastructural and operating costs, in a similar fashion as in the principle of economy of scale, a fundamental concept in economy that establishes the relationship between growth and production costs. This principle affirms that, as the quantity of produced units rises, the average cost per unit of production declines\cite{silberston1972economies}. This impacts the balance between the costs of producing and that of maintaining and operating the network. On one hand, this allows to simulate optimal networks that resemble those observed in real transportation systems more closely, as we tune $\beta$. On the other hand, by comparing the networks resulting for various values of $\beta$ with those observed from real data, we can also assess the impact of the two types of cost in the design of various urban infrastructures.   

We use this model to analyze several transportation networks from $18$ cities and a national rail network \cite{kujala2018collection, litvine2024, nyc_subway_lines}.
Despite the complex nature of the mechanisms driving the design of transportation networks,
we observe that multiple of the studied urban transportation networks follow a surprisingly similar topological pattern, as noticed in biological systems. We observed that in several cases, the optimal networks obtained with our approach have similar cost and performance to those observed in real ones.

\section*{RESULTS}\label{sec:results}

\subsection*{Modeling network design in transportation networks}\label{sec:model}

\begin{figure}[h!]
	\begin{minipage}{0.24\columnwidth}
		\centering
		\includegraphics[width=0.99\columnwidth]{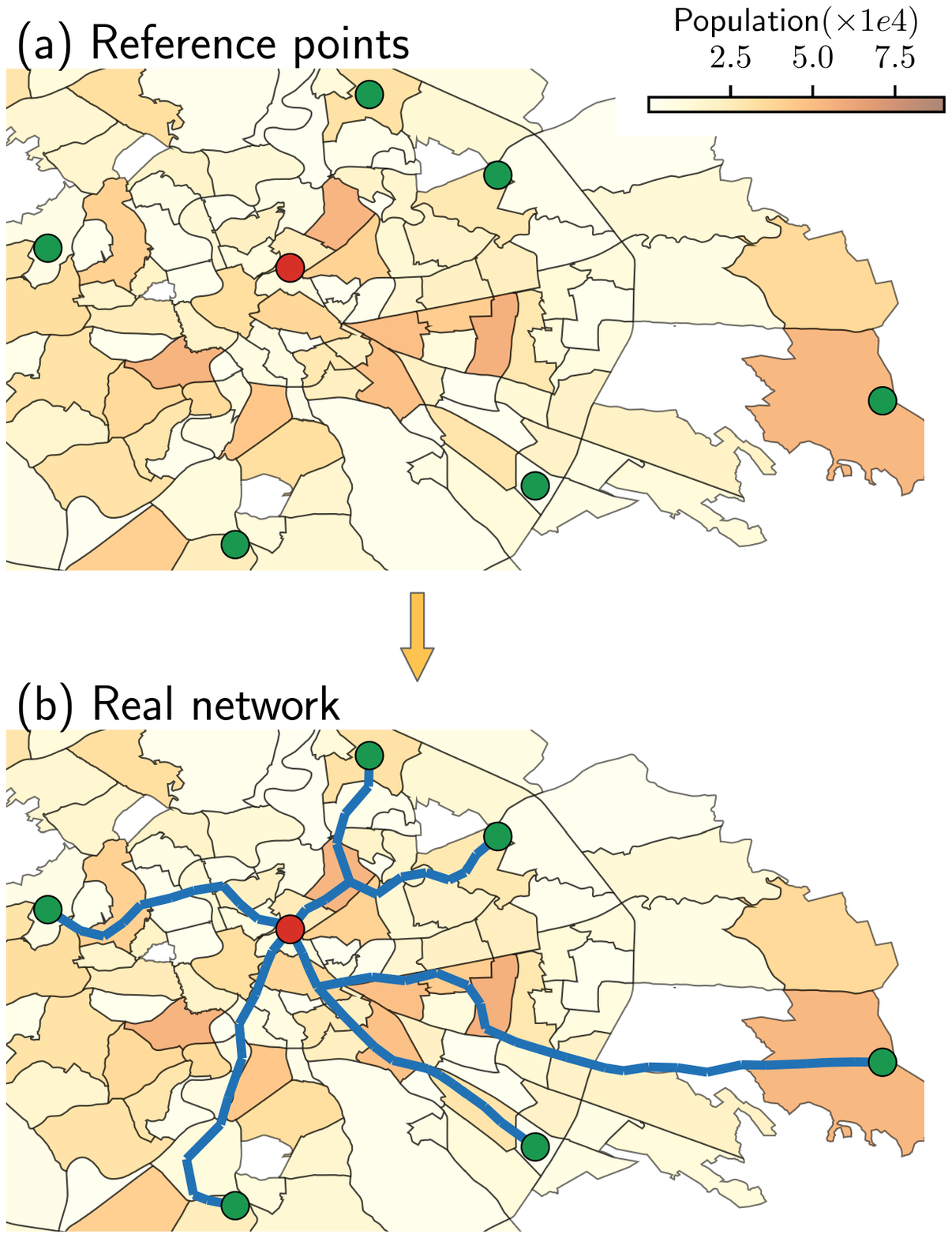}
	\end{minipage}
	\begin{minipage}{0.75\columnwidth}
		\centering
		\includegraphics[width=0.99\columnwidth]{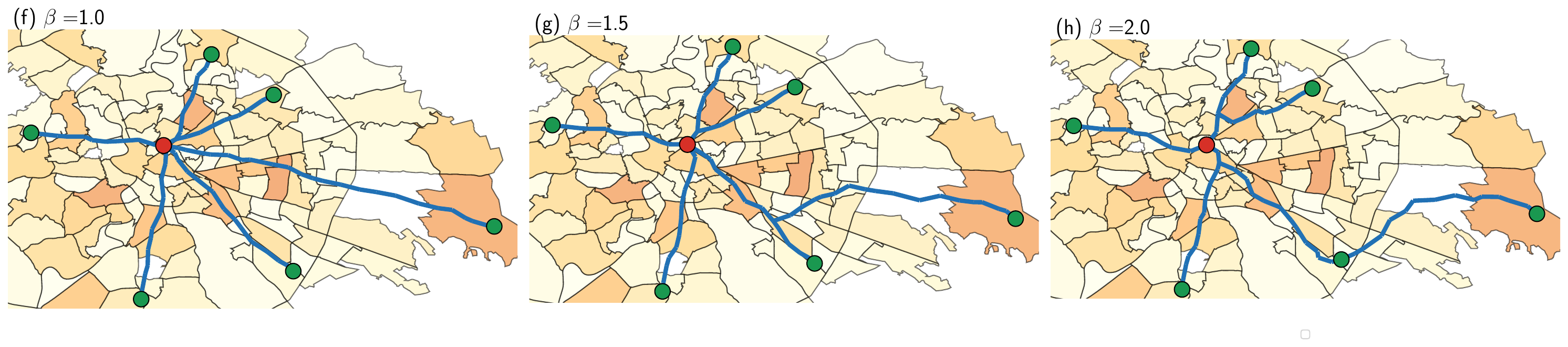}
		\\
		\includegraphics[width=0.99\columnwidth]{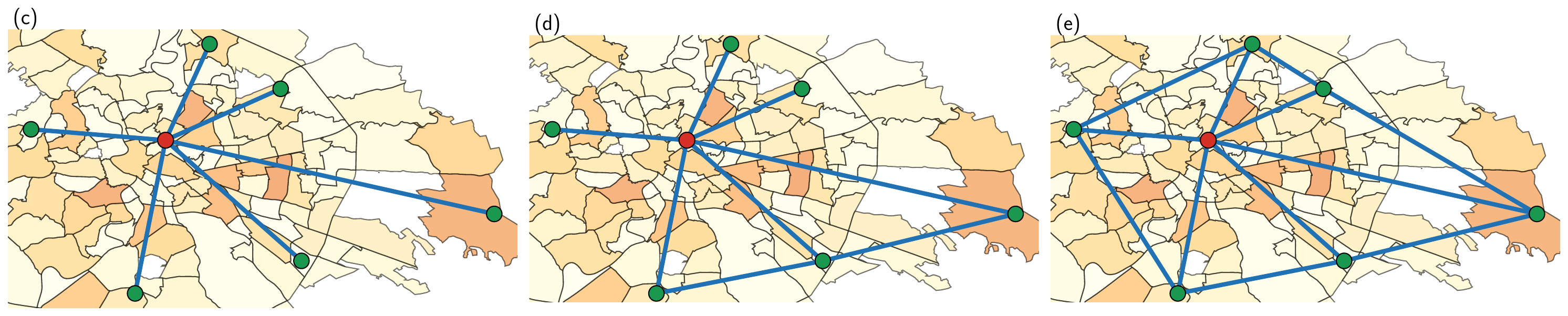}
	\end{minipage}
\caption{Problem setup for the subway network of Rome. (a) Given a set of real latitude-longitude coordinates denoting \source s (green) and \sink s (red), we aim to build a network structure that resembles well the observed public transportation network connecting those points, as in (b). (c)-(e) Intuitive ways to build a network structure by connecting \source s and \sink s, versus networks extracted with our optimal transport-based method in (f)-(h). The only known information is the set of six origins (O) and one destination (D). We capture different optimization mechanisms by tuning the $\beta$ parameter: in (f), the network is the shortest path-like structure, while in (g) and (h) we show examples of branched transportation schemes. This information is added to the population density across multiple urban areas (2019) \cite{populationdata}, where darker (lighter) colors indicate higher (lower) densities.}
\label{fig:problem-setup-rome}
\end{figure}

Consider an urban area where a set of points of interest are located in certain positions in space. These may correspond to a combination of central and peripheral points where people work and live. The goal is to connect them by building a transportation network under the perspectives of an optimality criterion, based on the minimization of a cost-based energy functional. At this point, we do not observe any network, but are rather free to use the whole space where the urban area is located, i.e. a 2D surface. From this, we need to select a set of points and edges connecting them, in other words, a network. In \Cref{fig:problem-setup-rome}a\textendash b we illustrate the problem setup for the subway network in Rome, where green and red markers denote an example set of such reference points and the lines denote edges in the observed metro network infrastructure. In the same figure, we show how existing stations are placed across urban areas with different population densities, as the evolution of subway networks often reflects the evolution of population and activity densities \cite{pei2022efficiency}.

In general, there are many choices for designing the network. For instance, in \Cref{fig:problem-setup-rome}c\textendash e we show three examples of intuitive shortest-path-like minimization solutions for the settings shown in \Cref{fig:problem-setup-rome}a. These are however quite different from the observed network in \Cref{fig:problem-setup-rome}b. The question we address is what network design principle is producing simulated networks that are more similar to those observed in real urban networks.
Optimality could be defined in various ways depending on the network engineers' and designers' goals, but generally, it is not known what principles they used when building the network. Instead, we want to assess this by observing real data of transportation networks and fitting them with a flexible and computationally efficient optimization setup guided by optimal transport theory. In this context, a well-defined cost-based functional combines aspects that are critical for a transportation network: the cost of building the infrastructure and that of operating the network, in terms of power dissipation. This is relevant in scenarios where we expect infrastructures to be regulated by energy-saving requirements and the principle of economy of scale, where it is more convenient to consolidate traffic into fewer and larger edges. We expect this to be a reasonable assumption in urban transportation networks. 
As with any other natural or urban system, we do not know \textit{a priori} what (if any) is the functional being optimized in the network under study. In fact, many of these systems (e.g., metro and tram) are built in phases \cite{derrible2010characterizing}, where the design of an initial backbone structure is followed by several expansion steps, which may lead to suboptimal structures. However, our model allows considering, among the many possible choices, a simple but yet principled mechanism of optimality. By measuring the degree of similarity of networks that follow these principles with existing urban transportation networks, we can assess if the observed ones can be explained by this simple mechanism. And if not, we can point out alternative infrastructures that can be better in terms of certain relevant network properties, e.g. total path length. Our setting is simple because we consider only a limited input (few nodes that need to exist, e.g. main origin and destination stations), but otherwise do not consider any other constraint, besides main physical laws such as conservation of mass, and let the model select nodes and edges from a two-dimensional space where it can be optimal to drive passengers, tuning only one parameter. 
For this, we adopt the formalism recently developed by Facca et al.\cite{facca2016towards,facca2019numerics,facca2020branch} that generalizes to a continuous space the original idea of Tero et al.\cite{tero2007mathematical}. In particular, this allows starting with only a set of relatively few \source \text{} and \sink \text{} nodes in input, and then designing a network by exploring the 2D surface, i.e. without the need of an initial existing backbone. The idea is inspired by the behavior of the slime mold \pp, which dynamically builds a network-like body shape when foraging. One can thus consider a dynamics for the two main quantities involved, flows and conductivities, that implements this mechanism at any point in space. The stationary solution of this dynamics corresponds to the minimizer of a Lyapunov cost in a standard optimization setup, which has a nice interpretation in terms of infrastructure and operating transportation costs.
From these solutions, one can then automatically extract optimal network structures using the approach presented in \cite{Nextrout}. From now onwards, we refer to the algorithmic implementation of this approach as ``Nextrout''.

Having introduced the main problem and ideas, we now briefly describe the model. Consider a surface in 2D and a set of points on it. Specifically, we denote a set of \source s\text{} and \sink s \text{}as $f^+$ and $f^-$, respectively. These contain the reference points where people enter and exit the transportation network. By defining $f = f^+ - f^-$, mass conservation can be enforced with the constraint $\int f dx = 0$. The two main quantities of interest are denoted with $\mu(x,t)$, the transport density (or conductivity), and $u(x,t)$ the transport potential. The former can be seen as a quantity proportional to the size of a network edge, while the latter determines the fluxes traveling along them. The dynamical equations in this continuous setting are

\begin{align}
-\nabla \cdot (\mu(t,x)\nabla u(t,x))  = f, 	\label{eq:dynamics-continuous1}\\
\frac{\partial \mu(t,x)}{\partial t}  = \left(\mu(t,x)\nabla u(t,x)\right)^{\beta} - \mu(t,x),	\label{eq:dynamics-continuous2}\\
\mu(0,x)  = \mu_0(x) > 0.	\label{eq:dynamics-continuous3}
\end{align}

Equation (1) determines the spatial balance of the flux, assumed to be governed by the Fick-Poiseuille flux as $q=-\mu\nabla u$; \Cref{eq:dynamics-continuous2} enforces optimal solutions, and represents the \pp {} dynamics in the continuous domain; \Cref{eq:dynamics-continuous3} is the initial condition. The parameter $\beta$ captures different optimization mechanisms: $\beta<1$ enforces congested transportation, $\beta=1$ is the shortest path-like and $\beta>1$ is branched transportation. In \Cref{fig:problem-setup-rome}f\textendash h we show examples of different optimal configurations, with $\beta=1$, $\beta=1.5$ and $\beta=2.0$.  Here, we consider the cases $1<\beta\leq 2$, where the approximate support of the conductivity $\mu$ displays a network-like structure. Under the lenses of a network, the conductivities can be viewed as the traffic capacities on the edges, hence \Cref{eq:dynamics-continuous3} defines how the initial traffic capacities are distributed along the network, while Eq. \Cref{eq:dynamics-continuous2} describes how these capacities evolve in response to the fluxes. As time evolves (i.e., $\lim_{t \rightarrow \infty}$), the equilibrium solution pair $(\mu^*,u^*)$ is reached. In \cite{facca2020numerical, facca2020branch} the authors show that under certain assumptions, this equilibrium solution is a minimizer of the functional
\begin{equation}\label{eqn:lyapunov}
\mathcal{L} (\mu,u) = \frac{1}{2}\int\mu|\nabla u|^2 dx + \int \frac{\beta}{2-\beta}\mu^{\frac{2-\beta}{\beta}} .
\end{equation}
This can be interpreted as the network transportation cost, where the first term is a network operating cost (or power dissipation, it is the Dirichlet energy to the solution of the first PDE), while the second is a non-linear cost to build the infrastructure. When $\beta>1$, this second term corresponds to a principle of economy of scale, where it is more convenient to consolidate traffic into fewer (but larger) edges. This is the scenario we consider here. By changing $\beta$, one can tune their relative contribution to the total transportation cost, thus tuning the impact of the principle of economy of scale and how much concentrated path trajectories are. Besides being relevant for urban transportation, this strategy seems to be a fundamental mechanism in various natural systems, e.g., tree branches and roots, blood vessels or river networks \cite{banavar1999size,banavar2014form}. \\
Alternative approaches can be considered to design a network infrastructure from simple mechanisms. Examples are cost-benefit analysis \cite{louf2013emergence}, maximizing for efficiency \cite{bontorin2023emergence} accounting for paths and flows of passengers, or minimizing the total length, as in the Euclidean minimum spanning tree problem. In discrete settings, when an initial network backbone is given, the cost in \Cref{eqn:lyapunov} has been shown to be implicitly related to the total path length minimization accounting for the passengers' trajectories \cite{lonardi2021designing}. One main difference between ours and these types of approaches is that we focus on a continuous space (as opposed to discrete settings) where the only necessary input is a set of origins and destinations, but otherwise no initial backbone network is given. This enables the design of a network from scratch, simulating where nodes and edges should be located in space to minimize the cost.

Once the optimal $(\mu^*,u^*)$ are obtained, one can then use the model described in \cite{Nextrout} to extract a final network structure, i.e. a set of nodes, a set of edges connecting them and their weights proportional to the conductivities. This can then be compared with the one observed from real data and repeated for various values of $\beta$. An example is shown in \Cref{fig:cost-spl-grenoble}a\textendash d, where, as we increase $\beta$, one can notice how the network infrastructure evolves from a shortest path-like ($\beta=1$) to a branching topology, where two branches ($\beta=1.5$) are created to lower the cost of building the infrastructure. In particular, the southern branch in \Cref{fig:cost-spl-grenoble}b resembles an analogous one observed in the real subway network of Rome. As $\beta$ increases to 2, this branch disappears to build a unique path that connects two destination points in the southeast side of the city, further lowering the cost to build the infrastructure. However, at this extreme value, the network is now less similar to the real one.\\

It is important to remark that in our setting, besides the parameter $\beta$, the other input quantities that need to be specified are origins and destinations via the function $f$. By imposing non-zero entries to this function, a user automatically selects a set of nodes that will be necessarily present in the output network. Otherwise, no other set of nodes or edges needs to be given, but is rather automatically learned by solving the optimization problem described above. This implies that similarity between simulated and observed networks trivially increases as we add more non-zero terms in $f$, as shown in \Cref{fig:cost-spl-grenoble}d. Here we consider the non-trivial scenario where we fix only a small number of origins and destinations, as described in more detail below.

\subsection*{Selecting Origin and Destination points}

\begin{figure}
	\centering
	\begin{subfigure}{0.51\textwidth}
		\includegraphics[width=0.99\textwidth,trim={0.1cm 0cm 0cm 0cm}, clip]{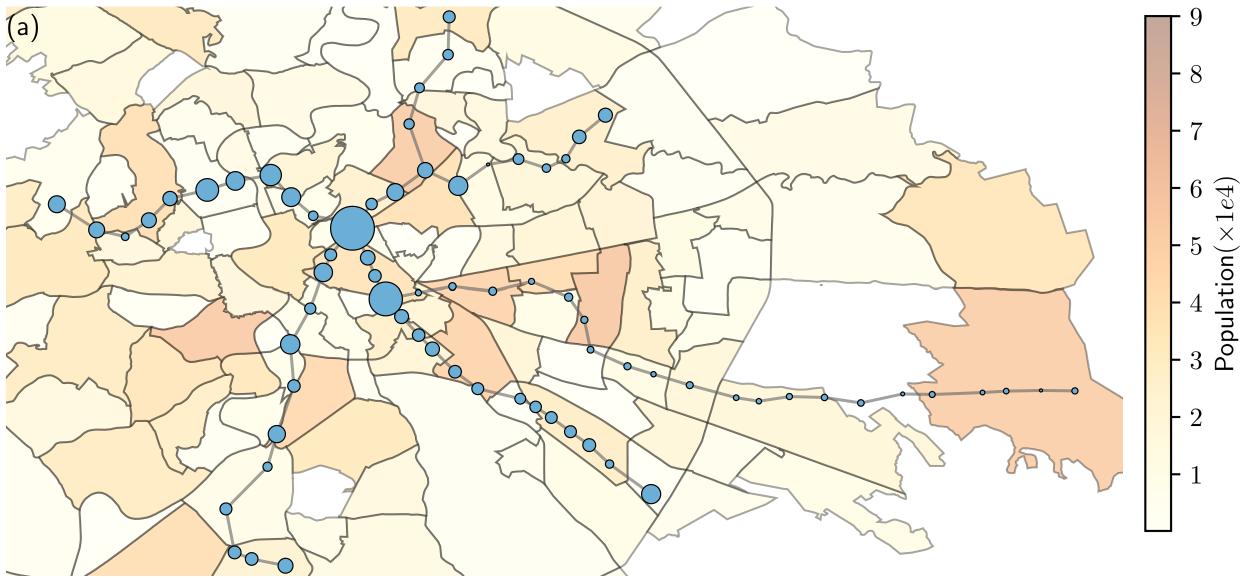}
	\end{subfigure}
	\begin{subfigure}{0.42\textwidth}
		\includegraphics[width=0.96\textwidth,trim={0cm 0cm 0cm 0cm}, clip]{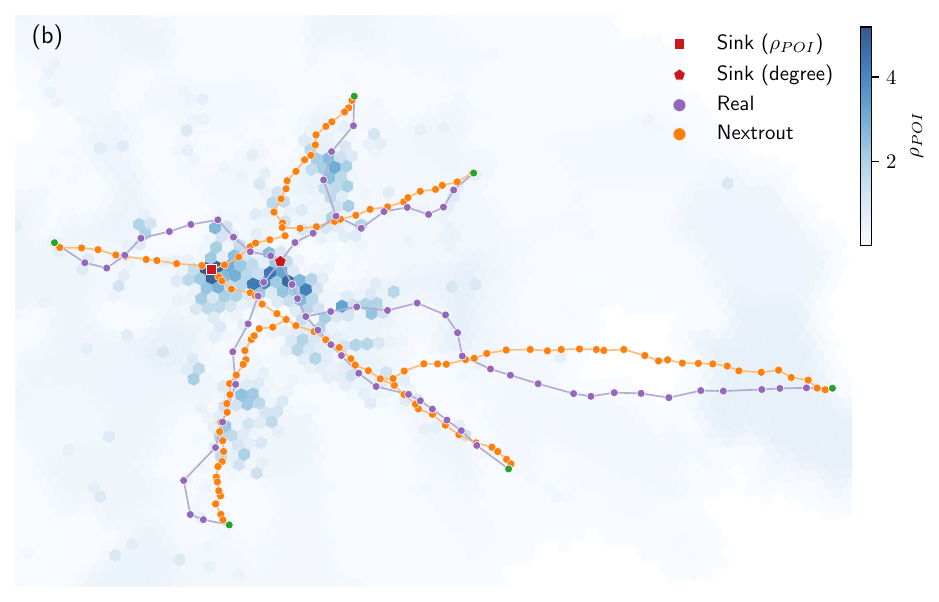}
	\end{subfigure}	
\caption{\textbf{Comparing criteria to select origin and destination nodes.} We compare two criteria to select the input nodes that we give to our algorithm, one based on a topological property (centrality) and one based on population and point-of-interest densities ($\rho_{\text{POI}}$). (a) Stations with the highest and lowest annual traffic (2019), proportional to the node sizes, over the population distribution in Rome (in multiples of 10000)). The degree centrality of nodes in the observed is related to the traffic at stations. In particular, the central node with the highest degree (destination), has also the highest annual traffic. (b) Density accounting for population and distribution of points-of-interest ($\rho_{\text{POI}}$), darker colors mean higher values, i.e. regions of higher relevance for transportation.
When accounting for the Points of Interest (POI), we notice that the city center presents higher density ($\rho_{\text{POI}}$) compared to the peripheral areas, therefore using centralities to select destinations is an approximation to real demands.} \label{fig:rome-pop-dense-poi}	
\end{figure}

As we aim at extracting a network, our problem starts by defining a set of \source s\text{} and \sink s (O-D) points in the space with coordinates $(x,y)$, where passengers might enter or exit. This choice necessarily impacts the output network, as the optimization problem depends on it. Ideally, one could reframe it by including O-D pairs as variables to be optimized along with conductivities and fluxes. But this becomes a different and more complex problem and it is not clear how to solve it. Here instead, we focus on the optimization setting introduced above and treat O-D pairs as fixed in input. While we limit selection to a small number of points, specifically one destination and few origins, it is important to decide where to place these input nodes in space. There are no universal criteria to define what are the most relevant points where city planners should add a stop to accommodate traffic when designing a network. However, existing infrastructures often evolve reflecting needs such as population increase or land usage \cite{yang2019integrated}. With this in mind, we can assume that at least some of the existing nodes have already been placed in positions relevant to transportation needs. Hence, we select O-D pairs from important nodes as observed in existing urban networks. Specifically, we use centrality measures obtained from the original network: nodes with the smallest and highest centrality are assigned as \source \text{} and \sink  \text{} nodes, respectively. These measures might reflect the choice on where to place new stations that are often made by urban planners or transportation engineers, usually based on a variety of factors, such as population density, land use patterns or available funding \cite{louf2014scaling}. For instance, in \Cref{fig:rome-pop-dense-poi}a we observe higher population density in peripheral regions, where the stations with lower centrality are located, whereas those with higher centrality are located towards the center, with lower population density. In the same figure, we show node sizes as proportional to the annual traffic in each station, as measured in 2019\cite{rome2019}. In this example, the highest traffic corresponds to the station with the highest degree centrality, thus reinforcing the choice of that node as a \sink.
	
Another possibility is to incorporate urban features by using some measure of attractiveness, which takes into account the densities of Points of Interest (POI) and population in a given urban area, an approach also used in other works \cite{bontorin2023emergence}. However, this strategy may not be scalable, as it requires the integration of several datasets, whereas using network centralities can be done automatically from the observed network data at no additional cost. We show an example of this for the metro network of Rome, to assess the extent to which these two strategies align. We notice that the high centrality nodes found by the two criteria are located nearby geographically and yield similar simulated networks, see \Cref{fig:rome-pop-dense-poi}. Hence, we adopt the centrality criteria as a good approximation for attractiveness to select origins and destinations in all the networks investigated here.

\subsection*{Investigating the similarity of optimal simulated networks and the observed transportation systems}
We apply the proposed dynamics to empirical data collected from $18$ different cities in multiple geographical regions around the world. For each city, we selected various available types of public transportation systems, such as rail, subway and tram, keeping the largest connected component. The networks considered in this manuscript have a few loops, as the dynamics can only retrieve loopless structures in the regime where network extraction is meaningful \cite{Nextrout}. 
These networks could be seen as phase I in the classification of \cite{derrible2010characterizing}, i.e. the initial phase where a backbone infrastructure is built, before a later evolution where further additional links are added through time. We expect these to be more likely to follow a global optimization criteria as the one formulated in our model (as opposed to other greedy heuristics for later extension phases).
We thus measure the loop ratio $L_{ratio}=n_{L}/E$ as the number of loops divided by the number of edges and select networks with a low ratio, i.e. with $L_{ratio}<0.2$ (see Methods for more details). One could in principle recover loopy structures by employing numerical schemes, e.g. superposition of different outputs \cite{baptista2021principled}, but this is not the main focus of this work. Instead, we point towards directions on the loop recover perspective, presented later in this manuscript, by exploring an example of a more complex network structure as the New York subway.\\

\begin{figure}[h]
	\centering
	\includegraphics[width=1.0\textwidth]{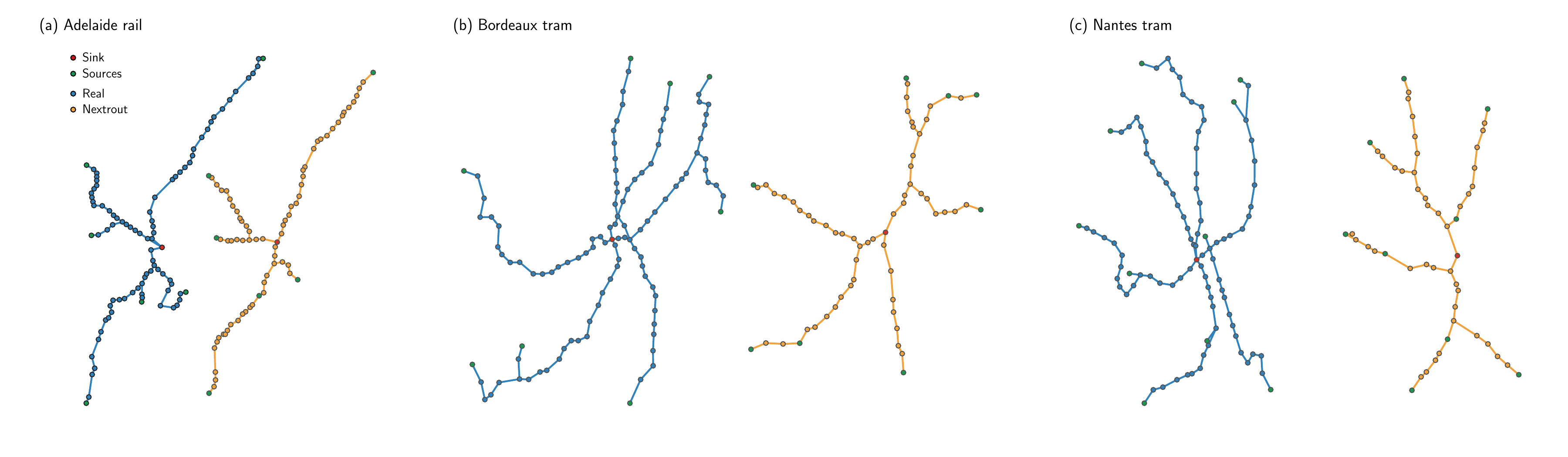}
	\caption{Example of different network topologies generated by \nextrout { }(yellow), plotted against the corresponding real network (blue). Green nodes represent those chosen as \source s{ }($\text{O}$), whilst red nodes correspond to the \sink s ($\text{D}$). (a) Adelaide rail network, with $N = 87$ nodes, $\text{O} = 6$ and $\text{D} = 1$. (b) Bordeaux tram network, with $N=108$ nodes, $\text{O} = 8$ and $\text{D} = 1$. (c) Nantes tram network with $N = 97$ nodes, $\text{O} = 10$ and $\text{D} = 1$.}
	\label{fig:examples}
\end{figure}

\begin{figure}[!h]
	\centering	
	\includegraphics[width=1.0\textwidth, clip]{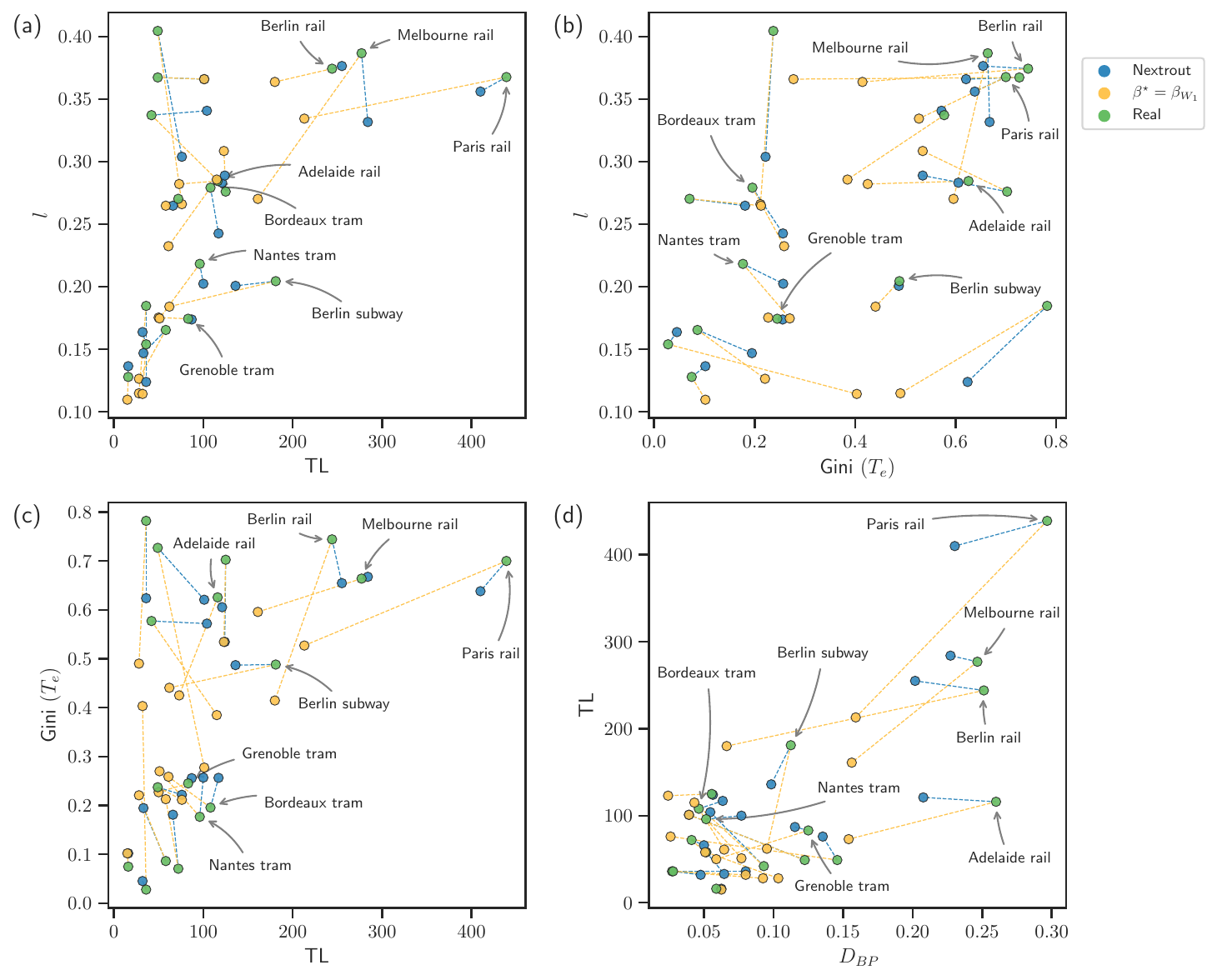}
	\caption{Performance measures for real and simulated networks. Each dataset is assigned to a different color, market shapes distinguish real and simulated networks. Simulated networks are further distinguished based on the one generated via \nextrout { } that gives the closest point in terms of the metrics plotted in the figure (circle) or the one corresponding to the best Wasserstein measure (square). (a) Cost (TL) measured in both simulated and real networks, plotted against the total path length. (b) Gini coefficient as a measure of traffic distribution, versus the total path length. (c) Traffic distribution in terms of the Cost. (d) Density of bifurcations plotted against the cost.}
	\label{fig:performance-measure}
\end{figure}

The applied Optimal Transport (OT) dynamics successfully describe the transportation network structures observed in different cities at a macroscopic level. While the selected transportation networks have different topologies and include multiple transportation modes, the networks reconstructed by Nextrout with only little information in input show a significant degree of similarity with the real ones (see \Cref{fig:examples}a\textendash c) for several of the studied cases, as evidenced by different similarity measures that we calculated to compare the topology of the simulated networks against the real ones, see \Cref{fig:performance-measure} and next sections for details. This suggests the existence of simple universal optimality rules captured by the so-called Dynamic Monge-Kantorovich (DMK) dynamics for the modeling of urban transportation rail networks, similar to what has been observed for the behavior of the slime mold \pp. Notice that in principle one can increase this similarity further, by simply adding more information in input in terms of origin and destinations (see \Cref{fig:cost-spl-grenoble}d for an example with multiple origins and destinations). However, here we are interested in recovering macroscopic structure in a more challenging scenario where input information is strictly limited to one central destination and few peripheral origins.

Beyond a qualitative visual comparison, we explore how our simulated networks score in terms of core network properties relevant for transportation compared to the real networks.
For this, we consider the cost, the total path length $l$, the distribution of traffic and the density of branching points (or bifurcations; here we use the two terms interchangeably), for both extracted and original networks. Similar to Tero\cite{tero2010rules}, we define the cost as the total length of the network (TL), i.e. the total number of edges. Passengers may not always take the shortest path, but may rather consolidate on fewer main arteries (e.g. to minimize the number of stops or connections)\cite{ibrahim2022sustainable}, a behavior that can be captured by a DMK discrete dynamics (built-in the filtering step of \nextrout) by varying $\beta$ and extracting the flows $u_{e}$ on edges, quantities proportional to the number of passengers using an edge. Hence, we consider an alternative measure of total path length as $l := \sum_{e\in E_i}\mathrm{l_e} |u_{e}|$, where $\mathrm{l_e}$ is the Euclidean distance. This takes into account $u_{e}$, the flow of passengers on an edge $e$, and its absolute value $|u_{e}|$ is proportional to the number of passengers traveling on an edge $e$, i.e. how traffic is distributed, assuming that passengers follow optimality principles to consolidate paths. This is a reasonable assumption in rail networks as the ones studied here, where the cost to build the infrastructure can be high (and thus should be minimized) and minimizing traffic congestion is not as relevant as in, e.g., road networks. In our experiments, we extract optimal flows $u_{e}$ by running the discrete DMK dynamics on the extracted and real networks, using the same sets of \source s and \sink s as used in the original network extraction problem, setting $\beta=1.5$. 
This information can also be used to measure the macroscopic behavior of traffic on edges, which can be measured using the Gini coefficient\cite{dixon1987bootstrapping} ($\text{Gini}(T_e)$) on the traffic $T_e = |u_{e}|$. This coefficient ranges from $[0,1]$, where the closer to $1$, the more unequal is the traffic distribution on the network. Finally, we calculate the percentage of bifurcations ($D_{BP}$) as the fraction of nodes with degree equal to $3$. 
In several cases, the simulated networks display similar properties as those observed on the real ones, as shown in \Cref{fig:performance-measure}. While similarity differs depending on the property and datasets vary in their range values, we notice that most of the datasets have at least a pair of properties that have a close value between simulated and observed networks. For instance, in \Cref{fig:examples}a we notice that Adelaide rail has an intuitively similar path length, which is confirmed in \Cref{fig:performance-measure}a. Furthermore, the same network shows comparable results for traffic and cost. As for the tram networks of Bordeaux and Nantes in \Cref{fig:examples}b-c, we observe similar behavior for cost and traffic.

\subsection*{Automatic selection of similar simulated networks}
Our method allows extracting various simulated networks by varying the parameter $\beta$. 
One can select the one that more closely resembles the observed one in terms of a particular metric of interest, as shown in the previous section. However, different metrics may lead to different most similar simulated networks (i.e. different $\beta$), which may not be ideal for a practitioner willing to consider an individual simulated network that resembles well the observed one in terms of all metrics. Hence, the need for a principle automatic selection criteria for choosing the value of  $\beta$. \\
The formalism introduced in the previous section suggests a natural way to tackle this problem by considering the Wasserstein similarity measure,  a main quantity in optimal transport theory \cite{santambrogio2015optimal}. Given two graphs that need to be compared, intuitively, this measure captures the minimum "effort" required to move a certain distribution of mass from one to the other. Similar ideas based on optimal transport to measure similarity between graphs have also been proposed in recent works \cite{maretic2019, xu2019gromov}. 
Here, we describe our proposal for a similarity measure and thus automatic selection of $\beta$ in detail.
Denote the observed network with $G_1(V_{1},E_{1})$ and the one obtained from the model introduced in previous sections with $G_2(V_{2},E_{2})$, where $V_{i},E_{i}$ denote the set of nodes and edges, respectively, $i=1,2$. 
We consider the union graph $G_U(V_{U},E_{U})$, with sets of nodes $V_{U} =V_{1} \cup V_{2}$ and edges $E_{U}=E_{1}\cup E_{2}$. One can further assign weights $w_{e}\in W_{U}$ to the edges $e \in E$, for instance using the  Euclidean distance $\ell_{e}$ between the nodes $i,j \in V_{U}$, where $e=(i,j)$, or simply binary values $\{0,1\}$. 
Notice that the observed network $G_{1}$ may contain nodes that do not correspond exactly to nodes in $G_{2}$, because in this continuous setting the model uses all the 2D space where the original network is embedded. Only the input \source \text{} and \sink \text{} nodes are guaranteed to be present in both graphs, as they are given in input to the model. 

\begin{figure}[!h]
	\centering
	\begin{subfigure}{1.0\textwidth}
		\includegraphics[width=1.0\textwidth,]{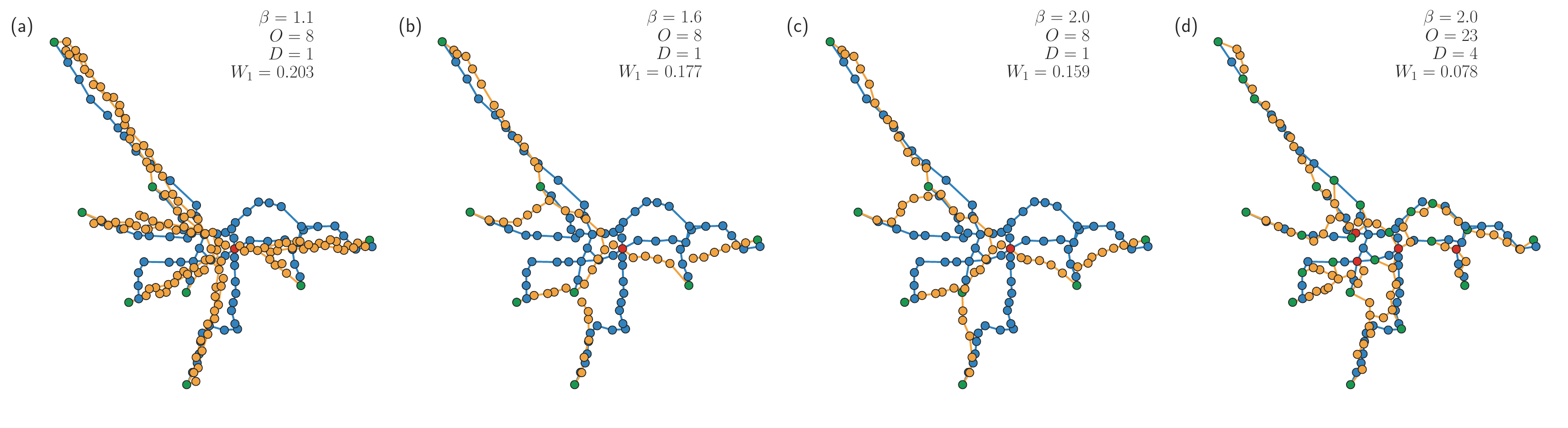}
	\end{subfigure}
	\begin{subfigure}{0.99\textwidth}
		\includegraphics[width=1.0\textwidth,clip]{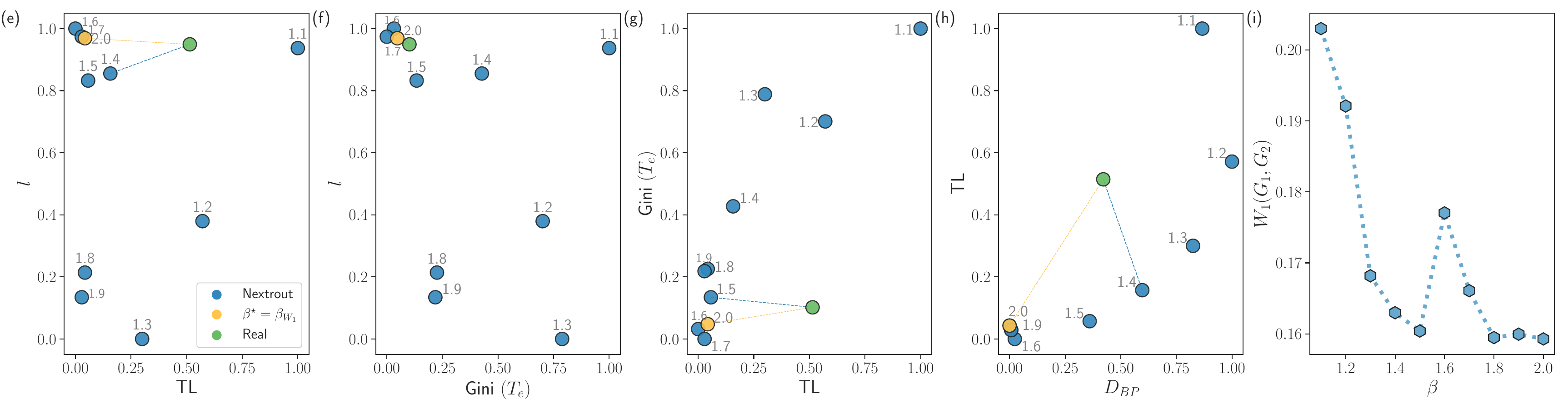} 
	\end{subfigure}	
	\caption{Wasserstein similarity measure between graphs for automatic selection of $\beta$. 
	(a-c): we select the origin nodes based on those with the smallest degree, and a unique destination as the one with the highest degree. 
	In (d) we show how the same network changes as we set more stations as initial input ($\text{O} = 23$ \source s and $\text{D} = 4$  \sink s), resulting in smaller Wasserstein, at the cost of a higher amount of information given in input.
	(e-h): we compare several network properties as measured in the observed and simulated networks. (e) The cost (TL) against the total path length ($l$), highlighting the different $\beta$ for each obtained network. 
	(f) $\text{Gini}(T_{e})$ against  $l$. The optimal network is equivalent to the one with minimal Wasserstein, i.e. $\beta=2.0$.		
	(g) TL against the Gini coefficient of traffic on edges ($\text{Gini}(T_{e})$). In this case, the optimal network corresponds to the one with $\beta=1.5$. 		
	(h) TL against the density of branching nodes. The closest network in terms of the number of bifurcations for $\beta=1.4$).
	(i) Wasserstein similarity measure for the simulated Grenoble tram networks as a function of $\beta$, in the setting of eight origins and one destination. The most similar network in terms of this measure is at $\beta=2$, when $W(G_1, G_2)$ is minimum. The peak at $\beta=1.6$ is due to the absence of a few edges in the rightmost part of the network that results in disconnecting a small branch, thus causing the distance to increase.}
	\label{fig:cost-spl-grenoble}
\end{figure}

Working on this union network, we then exploit a similar setting as the one already introduced with the model to obtain a Wasserstein-based similarity measure between $G_{1}$ and $G_{2}$. Specifically, we denote with $B$ the unsigned incidence matrix of $G_U$ with entries $B_{ie}=+1$ if node $i$ is a start or end point of the edge $e$, and $0$ otherwise.
Defining $q_{i}$ as an indicator vector for the edges in $G_{U}$ that are also in $G_{i}$, i.e. $q_{ie}=1$ if $e \in E_i$, and $q_{ie}=0$ otherwise, $\forall \ e \in  E_U$ and $i=1,2$, we can set the \source \text{} and \sink \text{} vectors $f = f^+-f^-$, such that $\boldsymbol{f^+} = B\,q_1$ and $\boldsymbol{f^-} = B\,q_2$, so that $G_{1}$ contributes to $f^+$ and $G_{2}$ to $f^-$. By running a discrete dynamics analogous to the continuous one described in \Crefrange{eq:dynamics-continuous1}{eq:dynamics-continuous3}, which can be done using \nextrout \cite{Nextrout}, one naturally obtains our Wasserstein similarity measure defined as:
\be\label{eqn:W}
W_{1}(G_{1},G_{2}) = \sum_{e\in E_{U}} w_{e}\,\mu_{e} \, ,
\ee
where $\mu_{e}$ are the optimal solutions for the conductivities on $G_{U}$ and $w_{e}$ is the weight of edge $e=(i,j)$. Here, we fix this to be the Euclidean distance between nodes $i$ and $j$. Examples of how the Wasserstein measure changes depending on the different output networks are shown in \Cref{fig:cost-spl-grenoble}i, where we show simulated networks and report their Wasserstein measure from the observed network of the Grenoble tram ($N=80$ nodes). Intuitively, the Wasserstein similarity captures how much ``cost'' is ``paid'' to move information between $G_1$ and $G_2$. This means that the more similar these networks are, the lower the Wasserstein is, i.e., when $G_1=G_2$, $W_1=0$, and if there are no nodes connecting them, $W_1 \rightarrow \infty$. In \Cref{fig:cost-spl-grenoble}b we show an example where $W_1$ is higher simply because there is a disruption in one of the branches of the network, thus increasing the cost to move from the real network to $G_2=G_{\beta=1.6}$. Notice that if similarity is chosen to be defined in terms of the cost, the closest network to the real one would be that with $\beta=1.4$, as shown in \Cref{fig:cost-spl-grenoble}e.\\
We further validate this measure by comparing it with other selection criteria based on the topological properties described above and found that the simulated graph selected with the Wasserstein measure has, on average, higher similarity with the real networks compared to the other selection criteria, across various properties. In other words, it shows transportation properties that are consistently more aligned to those behold by the observed network, see Supplementary Information S1.

\subsection*{The New York subway system: a look into more complex structures}\label{subsection:nycmetro}
The New York subway is one of the largest and busiest transportation systems in the world. Due to its size and complexity, navigating through such network might be difficult for humans\cite{gallotti2016lost}, but understanding its properties and structure could be  indicative of improvement to city planners and urban designers.
	
\begin{figure}
	\centering	
	\begin{subfigure}{0.36\textwidth}	
		\includegraphics[width=0.99\textwidth,trim={0cm 2.5cm 2.0cm 1.5cm}, clip]{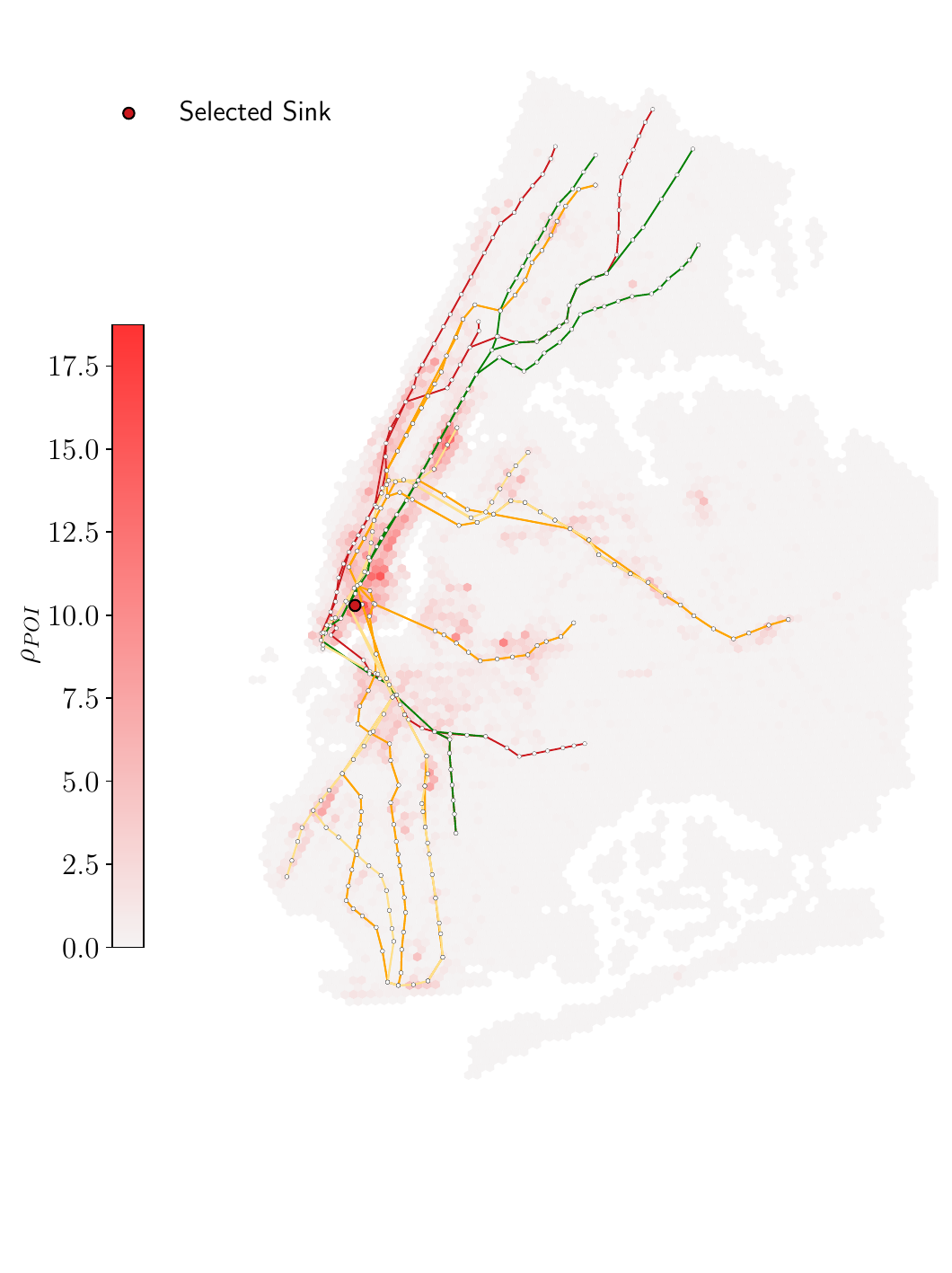}
	\end{subfigure}
	\begin{subfigure}{0.63\textwidth}
		\includegraphics[width=1.0\textwidth, clip]{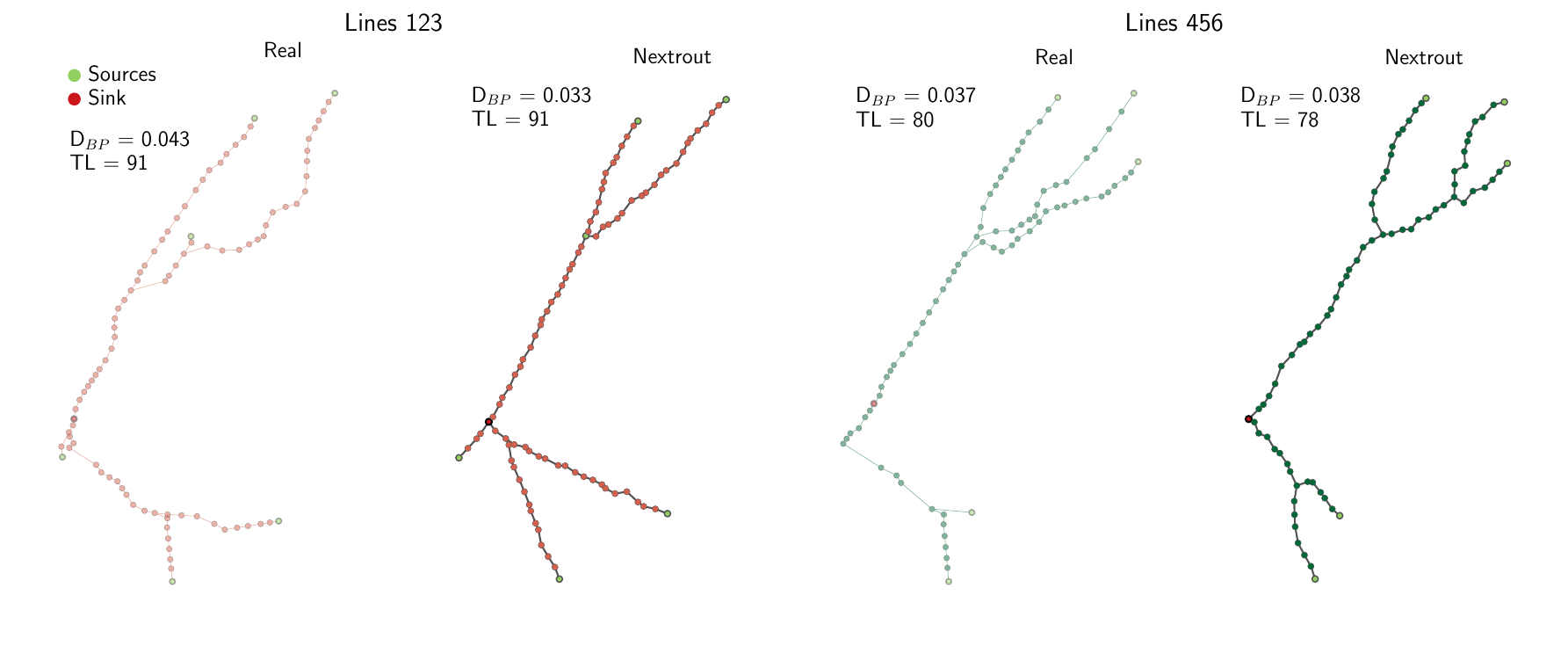}
		\includegraphics[width=1.0\textwidth, clip]{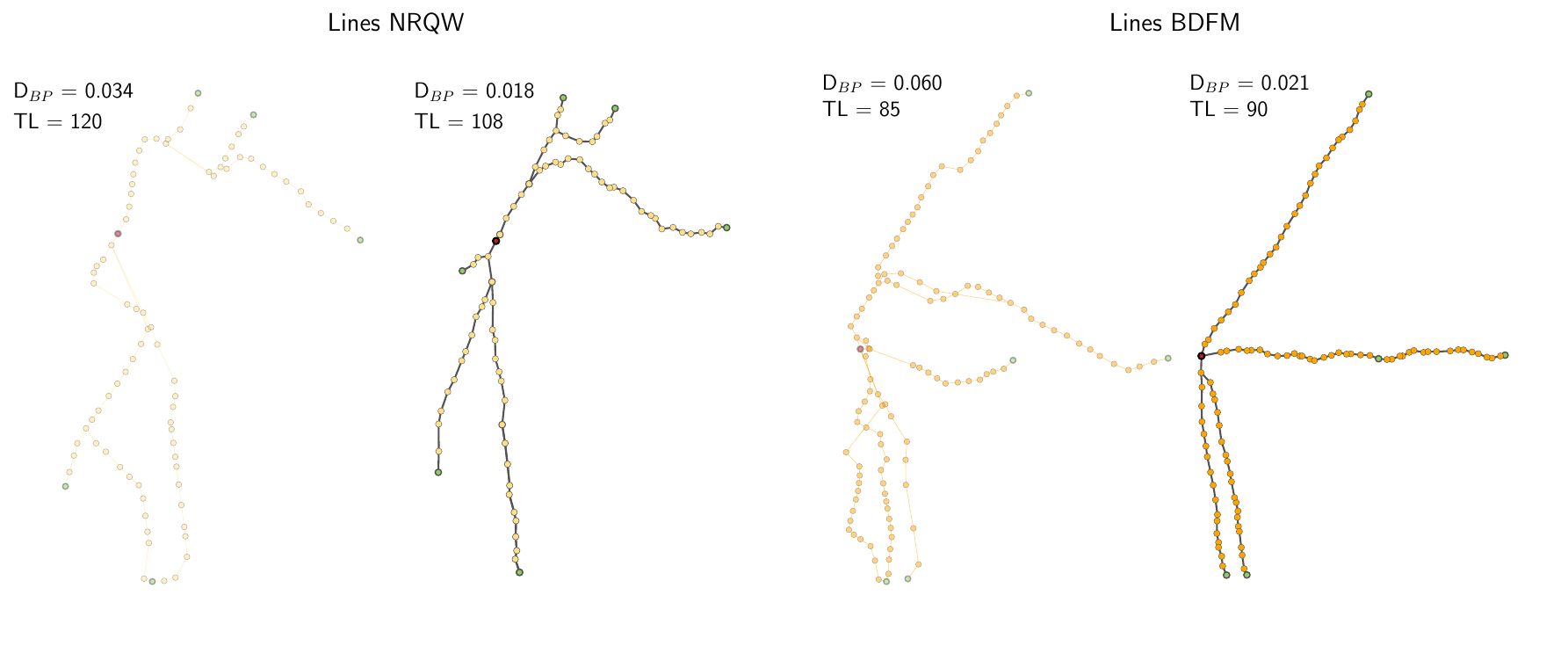}	
	\end{subfigure}

	\caption{Comparison for real and simulated networks for major lines of the subway system in New York, with distribution of POI. We select the four major lines (left panel), here shown along with the distribution of the density of POI (density varies as in the colorbar). We report the cost (TL) and density of branching points (D$_{BP}$) for both real and simulated networks. Here we set $\beta=1.1$ for the orange line, $\beta=1.5$ for both red and green lines, and $\beta=1.6$ for the yellow line.} \label{fig:NYC-poi-and-nextrout}
\end{figure}

In the scope of recovering such a complex structure, the strategy of selecting only a small number of \source s and \sink s, as done for the studied networks so far, might produce networks that are far from similar to the real one, especially given the high complexity of this particular system (see Fig. S7). We thus use a different approach that could be a pointer towards recovering structures from major urban transportation systems. Each line that comprises the subway infrastructure of New York could be seen as an independent network itself. With this assumption, we selected four major lines (red, green, orange and yellow), extracting the nodes with lower degree as \source s  and one common \sink for all of them, corresponding to the point with higher density of POI (see \Cref{fig:NYC-poi-and-nextrout}).

We notice that in terms of cost (TL), our simulated networks have similar values in all cases, with equal performance for the red line, and a small difference for all the other lines. For the density of branching points ($D_{BP}$) - besides an intuitive visual similarity - we observe comparable results.
For instance, the green line (456) has similar branches both on the north and south sides, and similar considerations apply to the red line (123) and the north side of the yellow line (NRQW). One can also investigate how main differences are distributed in the orange line (BDFM), where there is not much similarity between observed and simulated infrastructures. This is because two destinations on the east side are traversed by a unique branch in our simulated network, while the real one splits them into two branches. Furthermore, the southern part of the network also shows a distinct behavior, where our simulated network has two branches, the real one splits into a more complex pattern that cannot be explained by our optimality principles. While it is not clear whether these differences are due to different underlying optimization rules or a lack of optimality in the observed network, our method enables practitioners to identify key insights on principled alternative designs where optimality is clearly defined in terms of network operating and infrastructural costs.

\subsection*{Initial network development: the French Railway in the 1850s}
Our model builds a network backbone from scratch, with a global optimization that follows a principle of economy of scale. This could be particularly suited to study the initial development of a rail network infrastructure, as opposed to later stages where the network is gradually extended. We thus study a real scenario of the French railway system where we have access to historical information about network development in time \cite{litvine2024}, focusing on an initial phase around the year 1850. At that stage, the network topology contains multiple connected components but no loops - which only appear in a later stage, around several years later (see Supplementary Information S6, Fig. S9).

\begin{figure}[hbpt]
	\centering	
	\begin{subfigure}{0.31\textwidth}	
		\includegraphics[width=0.99\textwidth,trim={0.7cm 0.8cm 0.1cm 0.1cm}, clip]{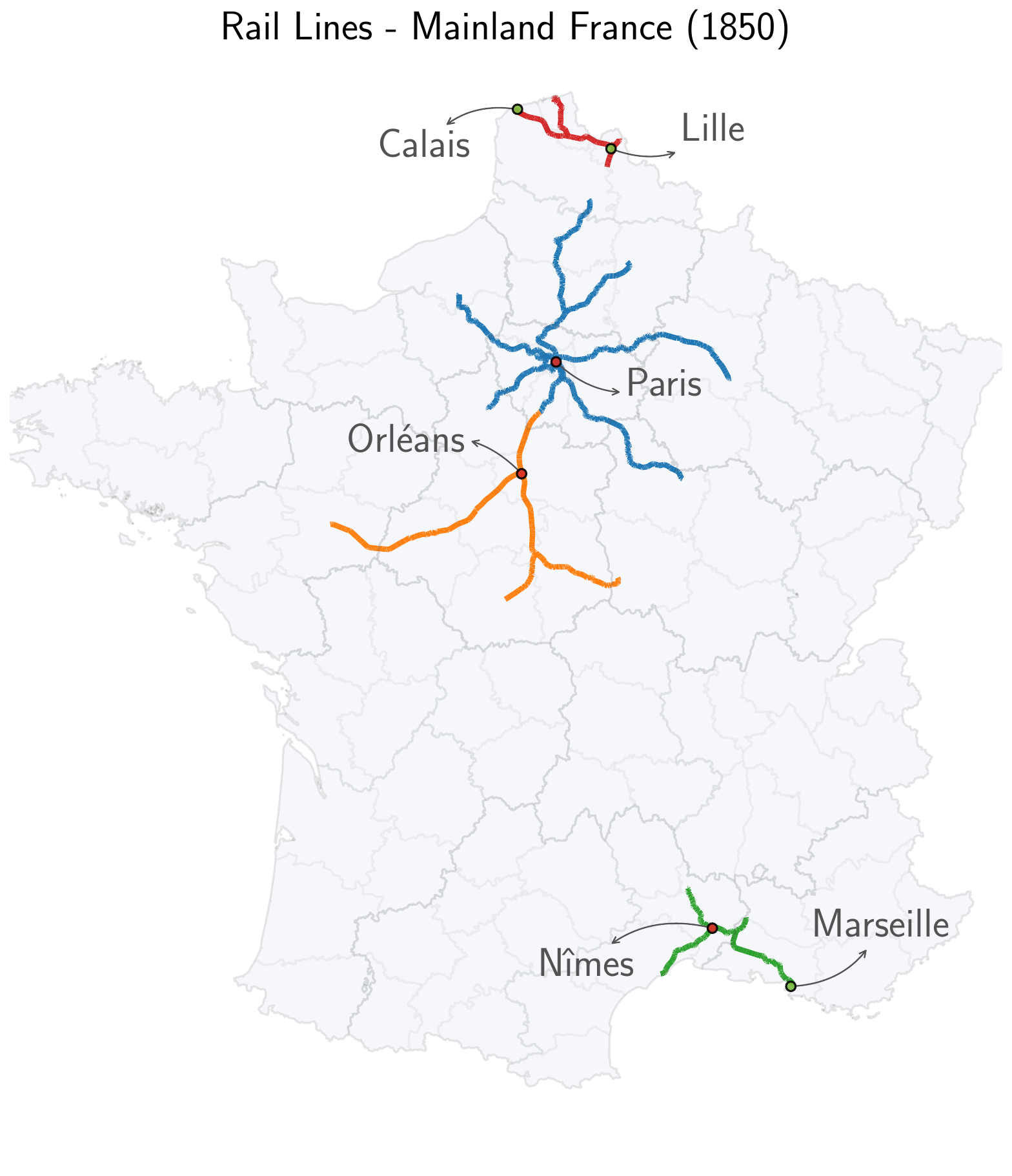}
	\end{subfigure}
	\begin{subfigure}{0.67\textwidth}
		\includegraphics[width=0.99\textwidth, clip]{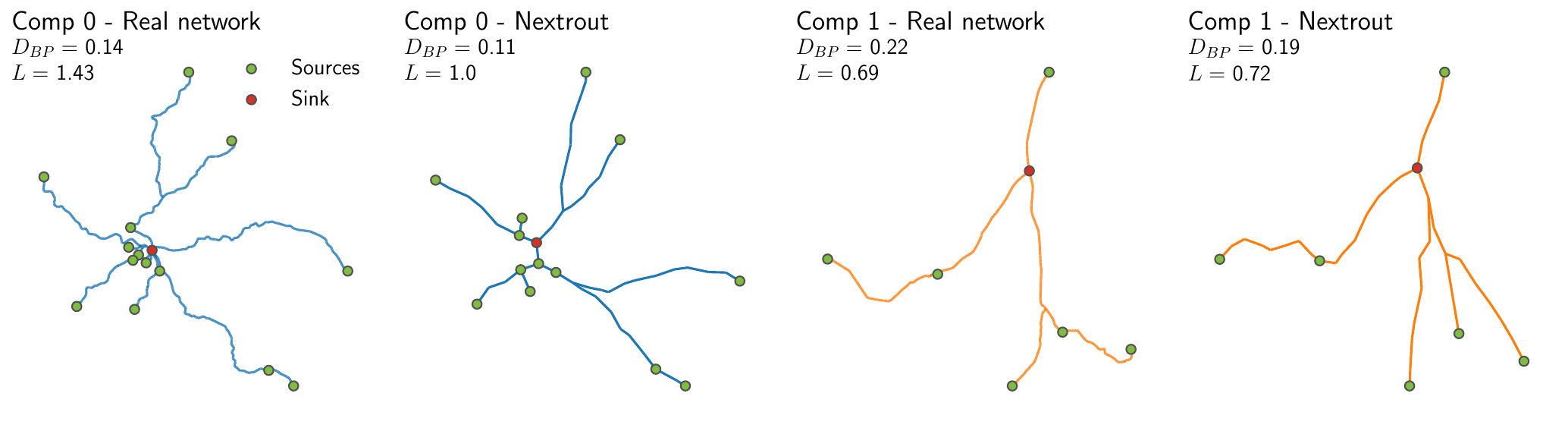}
		\includegraphics[width=1.0\textwidth, clip]{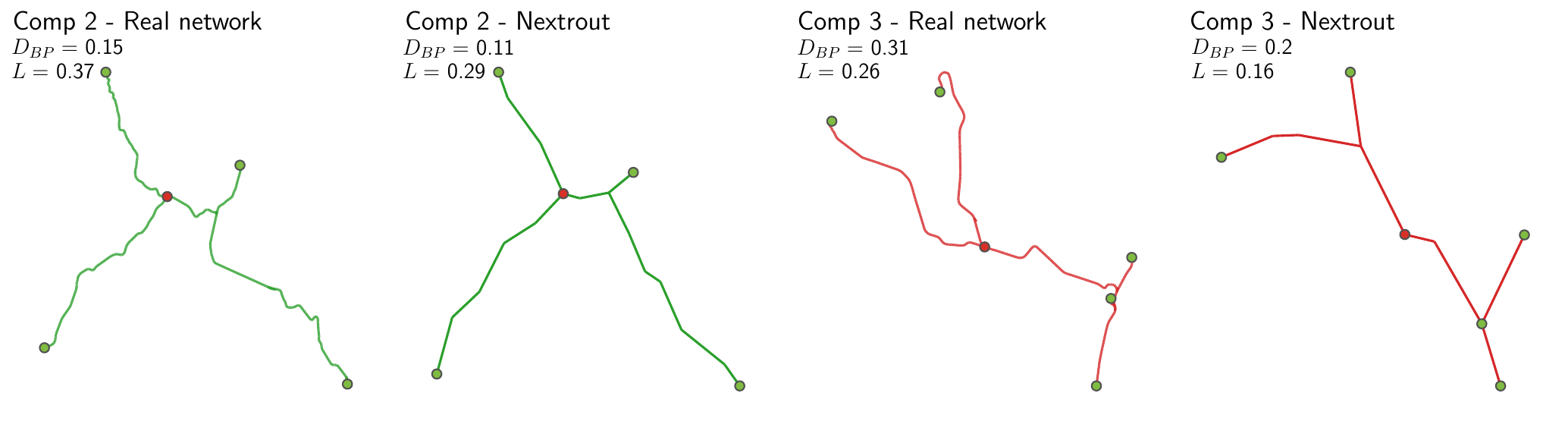}	
	\end{subfigure}
	
	\caption{Simulating the initial French Railways in the year 1850. Left: the original observed network, with connected components represented in different colors. Right: the networks simulated with our model for each component, given a set of \source s and one \sink. On top we report the density of branching points $D_{BP}$ and the total Euclidean length $L$.}
	\label{fig:FrenchRailways}
\end{figure}

We focus on the four biggest components with more interesting topologies, as the remaining components are either too small or simple straight lines, and select the node with the highest degree as a \sink. In the biggest component, for instance, this corresponds to Paris. In \Cref{fig:FrenchRailways} we show examples of real and simulated networks highlighting the $D_{BP}$ and the total length ($L$), the latter measured given the longitude and latitude coordinates mapped into a $[0,1]$ system of coordinates. We note various degrees of similarity in the different components between observed and simulated network. For instance, the biggest component has similar $D_{BP}$ but the observed network has a larger total length, mainly due to branches taking slightly longer detours to reach the sinks and the two small branches south-west of Paris being split from Paris onward into two in the observed network, while they are only later split in the simulated one. A similar behavior is observed in component 3 (north-est France), where we see the real network splitting earlier on than what simulated, thus causing a longer total length. The other two components have higher similarity in terms of both metrics, in particular, component 2 (south France) has a main branching point est of Nîmes similarly located in the observed and simulated network. The higher path length in this case is due to a longer detour of the southern branch. These types of detours could be caused by geographical obstacles that are not included in our more coarse-grained model. 

Understanding the underlying optimization principles that drive how transportation networks changed and evolved over time might point towards creating better transportation systems. Our approach is a pointer towards comprehending the initial stages of such evolution, particularly suited for systems that follow the principle of economy of scale.

\subsection*{Improving network properties} 
\begin{figure}[h]
	\centering
	\includegraphics[width=1.0\linewidth]{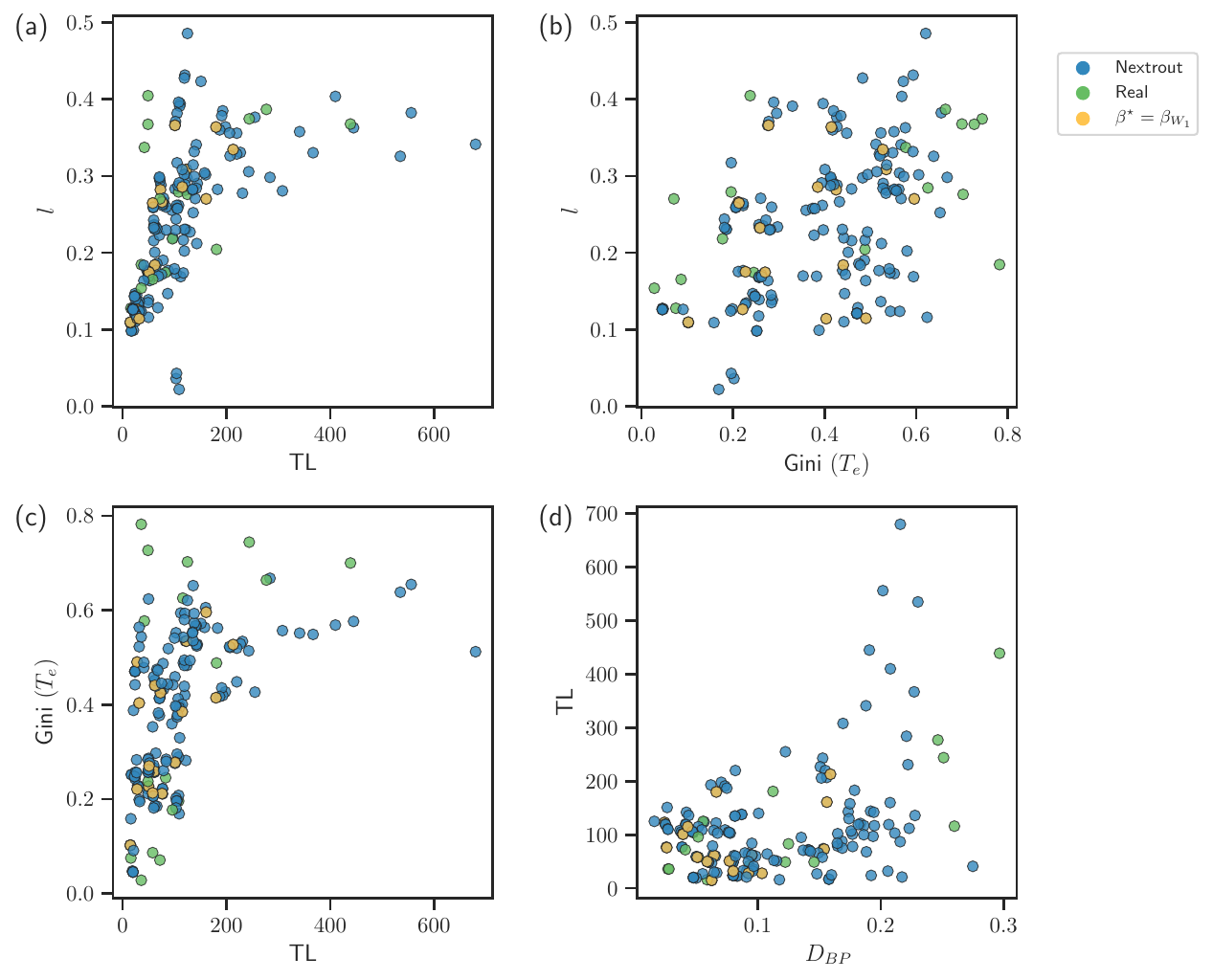} 
	\caption{Comparison of simulated and observed networks. We show the values of the main transportation properties investigated in this work for real and simulated networks. Simulated networks cover a wider range of properties' values, thus allowing in particular to select network that have lower or comparable values of these properties than those observed in the corresponding real networks. }
	\label{fig:all_betas_colors}	
\end{figure}

Simulating networks that follow optimality principles and resemble well those observed in real datasets can be used to assess how urban transportation networks perform in terms of main transportation properties. This can guide network managers towards potential measures directed at improving certain properties. This possibility is conveniently enabled by our approach, as by continuously tuning the parameter $\beta$ we can simulate various transportation scenarios, thus assessing how a network can increase or decrease a certain property. In \Cref{fig:all_betas_colors} we show the main properties in all the networks simulated with our model and compare with those observed in real data, aiming to compare their trade-offs between various performance metrics. In general, we notice how simulated networks cover a wider range of values than the real ones for the four transportation properties investigated in this work. This allows obtaining, for instance, networks that have shorter total path length $\ell$ with a comparable cost, as shown in  \Cref{fig:all_betas_colors}a where many simulated networks are located in the regime $0<\text{TL}<200$ with $\ell$ sharply dropping towards $0.1$, while many real networks have $\ell>0.1$. A similar behavior is observed also for the traffic against TL in \Cref{fig:all_betas_colors}c, where simulated networks cover areas of the plot where traffic is less congested (smaller $\text{Gini}(T_e)$), in contrast to several real networks. Among the simulated networks, those selected according to the best Wasserstein measure tend to have lower cost and a smaller percentage of bifurcations $D_{BL} $, indicating that this measure encourages not only the usage of a lower amount of edges, but also nodes with low degree.

\section*{Discussion}
Planning transportation systems in a city is a challenging task. This study has shown that some urban transportation systems with a small number of loops can be simulated by simple principles based on optimal transport theory and economy of scale.  
Using empirical data from one national railway and multiple rail transportation types across $18$ cities, our model provides simulated networks obtained with little information in input and no \textit{a priori} backbone network structure, exhibiting properties that resemble those observed in real transportation systems to various extents. 
Our model interpolates between various transportation regimes by tuning a single parameter, while allowing for a natural definition of a similarity measure to compare the simulated networks with those observed in real systems. We observed how the selected networks with this criterion can exhibit transportation properties that on average resemble the corresponding real networks well or point towards alternative infrastructures that improve relevant topological properties.\\

A limitation of this study is that our OT-based model does not capture infrastructures with loops, thus limiting its applicability to rail networks, or subway and rail networks with a very small density of loops. Possible extensions of the formalism presented in this work to account for loops are an interesting direction for future work. Besides simple heuristics and beyond the example we presented for the New York subway, one can make other interesting modeling choices to effectively tackle this problem. For instance, one could generalize our approach to situations where travel demands are treated stochastically \cite{corson2010fluctuations,hu2013adaptation,katifori2010damage} or change in time \cite{lonardi2021infrastructure}, scenarios where in certain regimes an OT-based approach can naturally lead to the formation of loops.
Similarly, loops could emerge in multicommodity settings where fluxes of passengers are distinguished by their origin and destination stations, using an OT-based multicommodity framework as the one proposed in \cite{lonardi2021designing,lonardi2021multicommodity,bonifaci2022physarum}.  Both directions, provided they could be generalized to a continuous setting as the one studied here, can potentially result in optimal simulated network infrastructures capturing properties that differ from the ones analyzed here, e.g. robustness to disruptions.\\
Another limitation is the assumption that networks are static, i.e. do not change in time. It would be interesting to compare how differences between simulated and observed networks may arise because different network branches may have been developed in different time periods. This could be used to study the evolution of transportation properties in time, as done in \cite{pei2022efficiency,cats2017topological}. Similarly, the networks studied in this work are often one layer of a multimode network. Integrating other transportation modes into a multilayer formalism and suitably adapting our OT-based approach, e.g. borrowing ideas from \cite{ibrahim2021optimal}, could give us a deeper understanding of optimal network design in interconnected urban systems.\\
Our model takes a few inputs (origins and destinations), but it does not consider any geographical obstacle. This can result in fine-grained mismatches between simulated and observed topologies due to longer detours in the real infrastructures to avoid these physical obstacles. In principle, this could be incorporated into our model by properly adding extra terms in the dynamical equations that drive flows and conductivities differently based on the location in space. However, this would require fine-grain details to be specified in input, an information that may not be easily available. 

In our work, we focus on designing a network from scratch. This is relevant in cases where infrastructures are relatively new or did not change considerably compared to their initial design, as in the cases we investigated here. However, this may be limited to studying infrastructures that have evolved over time. For these scenarios, it would be appropriate to consider how our model can be adapted to study network growth, where an initial backbone is further extended with new branches. This problem is related to the interplay between an urban transportation network and the distribution of its underlying population, as there could be a co-evolution between the two that should be taken into account \cite{brelsford2018toward}.

In summary, there are many factors contributing to the development of urban transportation networks. Our simple optimization scheme provides a principled and computationally efficient benchmark for comparison with real-world networks. By interpolating between different transportation regimes, we can vary the degree of similarity between the networks simulated by optimal transport principles and those observed in real systems. In particular, measuring relevant topological properties on simulated network resulting from different parameters' values against those observed in real networks can give us indications on how to improve transportation performance when taking into account principles of optimal transport and economy of scale.


\section*{Methods}\label{methods}

\subsubsection*{Data collection and analysis}

We collected network data from various public transportation networks from $18$ different cities \cite{kujala2018collection} and one national railway\cite{litvine2024}. Network statistics are detailed in \Cref{table:network_stats}. Each city had one or multiple transportation modes available. 
Node ids were associated with the longitude and latitude coordinates of real stations for multiple means of transportation, as well as possible connections between them. Our main goal was to analyze each network individually, therefore we did not address the multilayer case where connections among the different means of transportation exist. For instance, if rail and subway stops have the same coordinates, they are treated as distinct in each network.

To avoid possible redundancies or inconsistencies in the data, such as duplicated nodes or edges that looked too long, we performed a preprocessing step. Specifically, we considered a threshold $\tau$ that corresponds to the minimum distance in kilometers between the pairs of nodes that had the same node ids and no connections between them, matching features stored in the node metadata such as names of the real stations. If the Euclidean distance $d(i,j)$ between nodes $i$ and $j$ was smaller than this threshold, we collapsed the two nodes into one, i.e. $i=j$. This was used to avoid those entries and exits of each station would be counted as two distinct nodes in the same network, and possibly affecting the selection of \source s{ } and \sink s.

To match the latitude and longitude coordinates of the datasets with those in the $2$-dimensional plane that \nextrout \text{} uses to solve the continuous problem, we re-scaled every pair $(\text{lon},\text{lat})$ to a $(0,1)$ system of coordinates. Starting with a total of $64$ data points (networks), we extracted the individual disconnected components and the number of loops for each of them. Network extraction was performed on the biggest components only.

\begin{table*}[htpb]
\centering
\begin{tabular}{*{8}{l}}
	\toprule
	City & Transport mode &  $N$ &  $E$ & \# Comp &  $L_{ratio}$ & $O$ & $D_{\{d_i, B_i\}}$\\
	\midrule
	Adelaide & rail &  88 & 116 	& 1 & 0.25 &6 &1 \\
	Berlin & rail & 203 & 258 & 1  & 0.21 & 19 &1
	\\
	Berlin & subway &     169 &     181 &  1 &   0.072  &  14 & 1 \\
	Bordeaux &   tram &     110 &     110 & 1 &     0.009 &  8 & 1\\
	Brisbane &   rail &   297  &    367  &  1  & 0.193   & 10  & 1\\
	Dublin &   rail &    59  &    73  & 1&  0.03  &  10 & 1\\
	Grenoble &   tram &      80 &      83 & 1&    0.048 & 8 & 1\\
	Helsinki & subway &      17 &      16 &   1&   0.0 & 4& 1\\
	Lisbon &   rail &      48 &      49 &  2&  0.041 & 9& 1\\
	Luxembourg &  rail  &   43  & 56  & 1 &    0.025   &  11 & 1\\
	Melbourne &  rail  &  219  &  290  & 1  &   0.248  &  22 & 1\\
	Nantes &   tram &      97 &      96 &  1 &    0.0 & 10& 1\\
	New York (full) &   subway &      423 &      506 &  1 &    0.17 & 22& 1\\	
	Paris &  rail  &     337   & 445   & 1 &    0.244 & 24  & 1\\
	Prague & subway &      24 &      23 & 3 &     0.0 & 6& 1\\
	Rome & subway &      73 &      72 & 1 &     0.0 & 6& 1\\
	Toulouse & subway &      37 &      36 & 1 &     0.0 & 4& 1\\
	Venice &   tram &      37 &      38 & 1&     0.053 & 4& 1\\
	\bottomrule
	\end{tabular}
	\captionof{table}{Description of real networks considered. We report the main network statistics as number of nodes $N$, number of edges $E$, number of components \# Comp, number of origins $O$, of destinations $D$, and loops ratio $L_{ratio}$ defined as the number of loops divided by the number of edges, and selecting networks with $L_{ratio}<0.2$, as higher values would require the recovery of loops in the extracted networks.}\label{table:network_stats}

\end{table*}

We extract networks using \nextrout \text{} \cite{Nextrout}, selecting $1<\beta \leq 2$ such that for every pair $(\source s, \sink s)$ we simulate $10$ different networks. Since the extracted networks may contain redundancies, we remove them using the graph filtering step from \nextrout. Outputs of this step have less redundant structures and are closer to the optimal topologies. 

\paragraph{Selecting origins and destinations with points of interest.}
For the 16 studied networks, we used \source s and \sink s based on the degree and betweenness centrality measures. The degree centrality $d_{i}$ of a given node is defined as the number of edges connected to it. The betweenness centrality is defined as the frequency with which a node is on the shortest path between all other nodes,
$$
B_i = \sum_{i\neq j \neq k} \frac{\sigma_{ik}(j)}{\sigma_{ik}},
$$
where $\sigma_{ik}$ is the total number of shortest paths from node $i$ to node $k$ and $\sigma_{ik}(j)$ those shortest paths passing through $j$.

Terminals were chosen as follows: nodes with $d_{i}\leq1$ are assigned as \source s, while those with $d_{i}=\max_{n}\ccup{d_{n}}$ or $B_i = \max_{n}\ccup{B_{n}}$ as \sink s. We selected this set $\{\text{\source s,\sink s}\}$ to be small, in order to use the least amount of information in input. In multiple cases the set of \source s was equivalent in both centralities, with the difference being on the location of the \sink s, thus the output networks were different. In terms of final networks properties, the results are comparable for both studied centralities (see SI for more details).

In order to account for other measures of attractiveness for the \sink s and investigate the impact of including realistic information about urban areas, we explored a different approach for the cities of Rome and New York, by looking at a measure that accounts for both population distribution and land usage. To do that, we collected both population and distribution of POI from Open Street Map (OSM) data, and mapped them to an $\text{H}3$ tiling discretization of the space. We then defined the density of POI as $\rho_{\text{POI}} = P_{ij}W_{ij}$, where $P_{ij}$ is the population density and $W_{ij}$ is the number of POI for the corresponding $ij$ cell. Our hypothesis is that stations with higher centralities correspond to higher density cells. The \sink is then assigned based on the center of this H3 cell with highest $\rho_{\text{POI}}$. In \Cref{fig:rome-pop-dense-poi} (b) we show how this new criterion generates a configuration that leads to a different optimal network but still preserves similarity compared with the real network. We also notice that the highest centrality node is connected to regions with higher densities compared to the peripheral areas, where the \source s are placed.

\bibliographystyle{amsplain}
\bibliography{bibliography}

\section*{Acknowledgements}
The authors thank the International Max Planck Research School for Intelligent Systems (IMPRS-IS) for supporting Daniela Leite.

\section*{Data Availability}
The synthetic data ca be obtained from the corresponding author upon request.


\section*{Additional information}
\textbf{Accession codes}:  open source codes and executables are available at \href{https://github.com/Danielaleite/opt-urban-nextrout}{https://github.com/Danielaleite/opt-urban-nextrout}.\\
\textbf{Competing interests}. The authors declare no competing interests.

\end{document}